\documentclass[11pt]{article}

\title{Numerical Methods for Eigenvalue Distributions of Random Matrices}
\author{Alan Edelman and Per-Olof Persson}
\date{\today}

\usepackage{amsmath}
\usepackage{graphicx}

\begin{document}

\maketitle

\begin{abstract}
We present efficient numerical techniques for calculation of
eigenvalue distributions of random matrices in the beta-ensembles. We
compute histograms using direct simulations on very large matrices, by
using tridiagonal matrices with appropriate simplifications. The
distributions are also obtained by numerical solution of the
Painlev\'e II and V equations with high accuracy. For the spacings we
show a technique based on the Prolate matrix and Richardson
extrapolation, and we compare the distributions with the zeros of the
Riemann zeta function.
\end{abstract}

\section{Largest Eigenvalue Distributions}

In this section, the distributions of the largest eigenvalue of
matrices in the $\beta$-ensembles are studied. Histograms are created
first by simulation, then by solving the Painlev\'e II nonlinear
differential equation.

\subsection{Simulation}

The \emph{Gaussian Unitary Ensemble} (GUE) is defined as the Hermitian
$n\times n$ matrices $A$, where the diagonal elements $x_{jj}$ and the
upper triangular elements $x_{jk}=u_{jk}+iv_{jk}$ are independent
Gaussians with zero-mean, and
\begin{align}
\begin{cases}
\mathrm{Var}(x_{jj})=1, & 1\le j\le n, \\
\mathrm{Var}(u_{jk})=\mathrm{Var}(v_{jk})=\frac{1}{2}, & 1\le j<k\le n.
\end{cases}
\end{align}
Since a sum of Gaussians is a new Gaussian, an instance of these matrices
can be created conveniently in MATLAB:
\begin{small} \begin{verbatim}
  A=randn(n)+i*randn(n);
  A=(A+A')/2;
\end{verbatim} \end{small}
The largest eigenvalue of this matrix is about $2\sqrt{n}$. To get a
distribution that converges as $n\rightarrow\infty$, the shifted and
scaled largest eigenvalue $\lambda_\mathrm{max}'$ is calculated as
\begin{align}
\lambda_\mathrm{max}'=n^\frac{1}{6}\left(\lambda_\mathrm{max}-2\sqrt{n}\right).
\end{align}
It is now straight-forward to compute the distribution for
$\lambda_\mathrm{max}'$ by simulation:
\begin{small} \begin{verbatim}
  for ii=1:trials
    A=randn(n)+i*randn(n);
    A=(A+A')/2;
    lmax=max(eig(A));
    lmaxscaled=n^(1/6)*(lmax-2*sqrt(n));
    % Store lmax
  end

  % Create and plot histogram
\end{verbatim} \end{small}
The problem with this technique is that the computational requirements
and the memory requirements grow fast with $n$, which should be as
large as possible in order to be a good approximation of infinity.
Just storing the matrix $A$ requires $n^2$ double-precision numbers,
so on most computers today $n$ has to be less than $10^4$.
Furthermore, computing all the eigenvalues of a full Hermitian matrix
requires a computing time proportional to $n^3$. This means that it
will take many days to create a smooth histogram by simulation, even
for relatively small values of $n$.

To improve upon this situation, another matrix can be studied that has
the same eigenvalue distribution as $A$ above. In \cite{AlanIoana}, it was
shown that this is true for the following \emph{symmetric} matrix when
$\beta=2$:
\begin{align} \label{eq:Hbeta}
H_\beta \sim \frac{1}{\sqrt{2}}
\begin{pmatrix}
N(0,2) & \chi_{(n-1)\beta} & & & \\
\chi_{(n-1)\beta} & N(0,2) & \chi_{(n-2)\beta} & & \\
 & \ddots & \ddots & \ddots & \\
 & & \chi_{2\beta} & N(0,2) & \chi_\beta \\
 & & & \chi_\beta & N(0,2)
\end{pmatrix}.
\end{align}
Here, $N(0,2)$ is a zero-mean Gaussian with variance $2$, and $\chi_d$
is the square-root of a $\chi^2$ distributed number with $d$ degrees
of freedom. Note that the matrix is symmetric, so the subdiagonal and
the superdiagonal are always equal.

This matrix has a tridiagonal sparsity structure, and only $2n$
double-precision numbers are required to store an instance of it. The
time for computing the largest eigenvalue is proportional to $n$,
either using Krylov subspace based methods or the method of bisection
\cite{Trefethen}.

In MATLAB, the built-in function \texttt{eigs} can be used, although
that requires dealing with the sparse matrix structure. There is also
a large amount of overhead in this function, which results in a
relatively poor performance. Instead, the function \texttt{maxeig} is
used below to compute the eigenvalues. This is not a built-in function
in MATLAB, but it can be downloaded from
\textsf{http://www-math.mit.edu/$\sim$persson/mltrid}. It is based on the
method of bisection, and requires just two ordinary MATLAB vectors as
input, corresponding to the diagonal and the subdiagonal.

It also turns out that only the first $10n^\frac{1}{3}$ components of
the eigenvector corresponding to the largest eigenvalue are
significantly greater than zero. Therefore, the upper-left
$n_\mathrm{cutoff}$ by $n_\mathrm{cutoff}$ submatrix has the same
largest eigenvalue (or at least very close), where
\begin{align}
n_\mathrm{cutoff}\approx 10 n^\frac{1}{3}.
\end{align}
Matrices of size $n=10^{12}$ can then easily be used since the
computations can be done on a matrix of size only
$10n^\frac{1}{3}=10^5$. Also, for these large values of $n$ the
approximation $\chi^2_n\approx n$ is accurate.

A histogram of the distribution for $n=10^9$ can now be created using
the code below.

\begin{small} \begin{verbatim}
  n=1e9;
  nrep=1e4;
  beta=2;
  
  cutoff=round(10*n^(1/3));
  d1=sqrt(n-1:-1:n+1-cutoff)'/2/sqrt(n);
  
  ls=zeros(1,nrep);
  for ii=1:nrep
    d0=randn(cutoff,1)/sqrt(n*beta);
    ls(ii)=maxeig(d0,d1);
  end
  
  ls=(ls-1)*n^(2/3)*2;
  
  histdistr(ls,-7:0.2:3)
\end{verbatim} \end{small}
where the function \texttt{histdistr} below is used to histogram the data.
It assumes that the histogram boxes are equidistant.
\begin{small} \begin{verbatim}
  function [xmid,H]=histdistr(ls,x)
  
  dx=x(2)-x(1);
  H=histc(ls,x);
  H=H(1:end-1);
  H=H/sum(H)/dx;
  xmid=(x(1:end-1)+x(2:end))/2;
  
  bar(xmid,H)
  grid on
\end{verbatim} \end{small}

\begin{figure}
  \begin{center}
    \includegraphics[width=\textwidth]{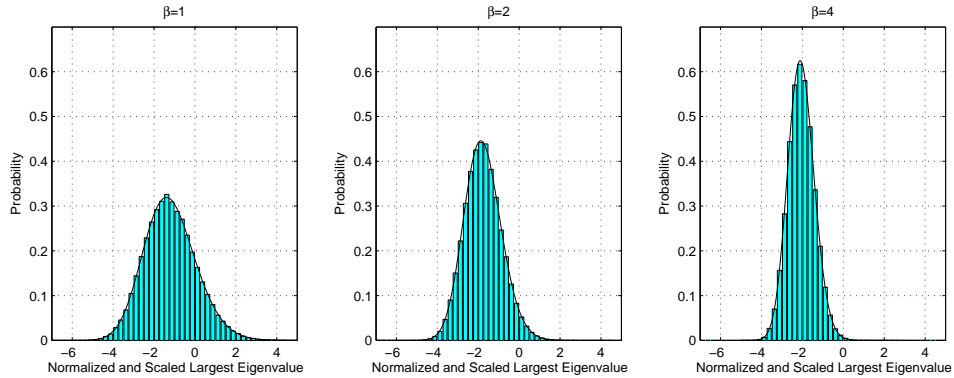}
  \end{center}
  \caption{Probability distribution of scaled largest eigenvalue ($10^5$ repetitions, $n=10^9$)} 
  \label{fig:plt1}
\end{figure}

The resulting distribution is shown in Figure~\ref{fig:plt1}, together
with distributions for $\beta=1$ and $\beta=4$. The plots also contain
solid curves representing the ``true solutions'' (see next section).

\subsection{Painlev\'e II}

Instead of using simulation to plot the distributions of the largest
eigenvalues, it can be computed from the solution of the Painlev\'e II
nonlinear differential equation \cite{TracyWidomLargest}:
\begin{align} \label{eq:painleve}
q''=sq+2q^3
\end{align}
with the boundary condition
\begin{align} \label{eq:bndcond}
q(s)\sim \mathrm{Ai}(s),
   \qquad\textrm{as }s\rightarrow \infty.
\end{align}
The probability distribution $f_2(s)$, corresponding to $\beta=2$, is then
given by
\begin{align} \label{eq:f2}
f_2(s)=\frac{d}{ds}F_2(s),
\end{align}
where
\begin{align} \label{eq:F2}
F_2(s)=\exp \left( -\int_s^\infty (x-s)q(x)^2\,dx \right).
\end{align}
The distributions for $\beta=1$ and $\beta=4$ are the
derivatives of
\begin{align}
F_1(s)^2 &= F_2(s)e^{-\int_s^\infty q(x)\,dx} \label{eq:F1} \\
\intertext{and}
F_4\left(\frac{s}{2^\frac{2}{3}}\right)^2 &=
  F_2(s)\left(\frac{e^{\frac{1}{2}\int_s^\infty q(x)\,dx}+
                    e^{-\frac{1}{2}\int_s^\infty q(x)\,dx}}{2} \label{eq:F4}
      \right)^2.
\end{align}
To solve this numerically using MATLAB, first rewrite~(\ref{eq:painleve})
as a first-order system:
\begin{align}
\frac{d}{ds}
\begin{pmatrix}
q \\ q'
\end{pmatrix}
=
\begin{pmatrix}
q' \\ sq+2q^3
\end{pmatrix}.
\end{align}
This can be solved as an initial-value problem starting at $s=s_0=$
sufficiently large positive number, and integrating along
the negative $s$-axis. The boundary condition~(\ref{eq:bndcond}) then
becomes the initial values
\begin{align}
\left\{
\begin{array}{rcl}
q(s_0) &=& \mathrm{Ai}(s_0) \\
q'(s_0) &=& \mathrm{Ai}'(s_0).
\end{array}
\right.
\end{align}

Although the distributions can be computed from $q(s)$ as a
post-processing step, it is most convenient to add a few variables and
equations to the ODE system. When computing $F_2(s)$, the quantity
$I(s)=\int_s^\infty (x-s)q(x)^2\,dx$ is required. Differentiate this
twice to get
\begin{align}
I'(s) &= -\int_s^\infty q(x)^2\,dx \\
\intertext{and}
I''(s) &= q(s)^2.
\end{align}
Add these equations and the variables $I(s),I'(s)$ to the ODE system,
and the solver will do the integration. This is not only easier and
gives less code, it will also give a much more accurate solution since
the same tolerance requirements are imposed on $I(s)$ as on the
solution $q(s)$.

In a similar way, the quantity $J(s)=\int_s^\infty q(x)\,dx$ is needed
when computing $F_1(s)$ and $F_4(s)$. This is handled by adding the
variable $J(s)$ and the equation $J'(s)=-q(s)$.

The final system now has the form
\begin{align} \label{eq:final}
\frac{d}{ds}
\begin{pmatrix}
q \\ q' \\ I \\ I' \\ J
\end{pmatrix}
=
\begin{pmatrix}
q' \\ sq+2q^3 \\ I' \\ q^2 \\ -q
\end{pmatrix}
\end{align}
with the initial condition
\begin{align} \label{eq:finalinitial}
\begin{pmatrix}
q(s_0) \\ q'(s_0) \\ I(s_0) \\ I'(s_0) \\ J(s_0)
\end{pmatrix}
=
\begin{pmatrix}
\mathrm{Ai}(s_0) \\ \mathrm{Ai}'(s_0) \\ \int_{s_0}^\infty(x-s_0)\mathrm{Ai}(x)^2\,dx \\
\mathrm{Ai}(s_0)^2 \\ \int_{s_0}^\infty\mathrm{Ai}(x)\,dx
\end{pmatrix}.
\end{align}
This problem can be solved in just a few lines of MATLAB code using
the built-in Runge-Kutta based ODE solver \texttt{ode45}.
First define the system of equations as an inline function
\begin{small} \begin{verbatim}
  deq=inline('[y(2); s*y(1)+2*y(1)^3; y(4); y(1)^2; -y(1)]','s','y');
\end{verbatim} \end{small}
Next specify the integration interval and the desired output times.
\begin{small} \begin{verbatim}
  s0=5;
  sn=-8;
  sspan=linspace(s0,sn,1000);
\end{verbatim} \end{small}
The initial values can be computed as
\begin{small} \begin{verbatim}
  y0=[airy(s0); airy(1,s0); ...
      quadl(inline('(x-s0).*airy(x).^2','x','s0'),s0,20,1e-25,0,s0); ...
      airy(s0)^2; quadl(inline('airy(x)'),s0,20,1e-18)];
\end{verbatim} \end{small}
where the \texttt{quadl} function is used to numerically approximate
the integrals in~(\ref{eq:finalinitial}). Now, the integration
tolerances can be set and the system integrated:
\begin{small} \begin{verbatim}
  opts=odeset('reltol',1e-13,'abstol',1e-15);
  [s,y]=ode45(deq,sspan,y0,opts);
\end{verbatim} \end{small}
The five dependent variables are now in the columns of the MATLAB
variable \texttt{y}. Using these, $F_2(s),F_1(s)$, and $F_4(s)$ become
\begin{small} \begin{verbatim}
  F2=exp(-y(:,3));
  F1=sqrt(F2.*exp(-y(:,5)));
  F4=sqrt(F2).*(exp(y(:,5)/2)+exp(-y(:,5)/2))/2;
  s4=s/2^(2/3);
\end{verbatim} \end{small}
The probability distributions $f_2(s),f_1(s)$, and $f_4(s)$ could be
computed by numerical differentiation:
\begin{small} \begin{verbatim}
  f2=gradient(F2,s);
  f1=gradient(F1,s);
  f4=gradient(F4,s4);
\end{verbatim} \end{small}
but it is more accurate to first do the differentiation symbolically:
\begin{align}
f_2(s) &= -I'(s)F_2(s) \\
f_1(s) &= \frac{1}{2F_1(s)}
  \left( f_2(s)+q(s)F_2(s) \right) e^{-J(s)} \\
f_4(s) &= \frac{1}{2^\frac{1}{3}4F_4(s)}
  \left( f_2(s)\left(2+e^{J(s)}+e^{-J(s)}\right)+
         F_2(s)q(s)\left(e^{-J(s)}-e^{J(s)}\right) \right)
\end{align}
and evaluate these expressions:
\begin{small} \begin{verbatim}
  f2=-y(:,4).*F2;
  f1=1/2./F1.*(f2+y(:,1).*F2).*exp(-y(:,5));
  f4=1/2^(1/3)/4./F4.*(f2.*(2+exp(y(:,5))+exp(-y(:,5)))+ ...
                       F2.*y(:,1).*(exp(-y(:,5))-exp(y(:,5))));
\end{verbatim} \end{small}

\begin{figure}
  \begin{center}
    \includegraphics[width=\textwidth]{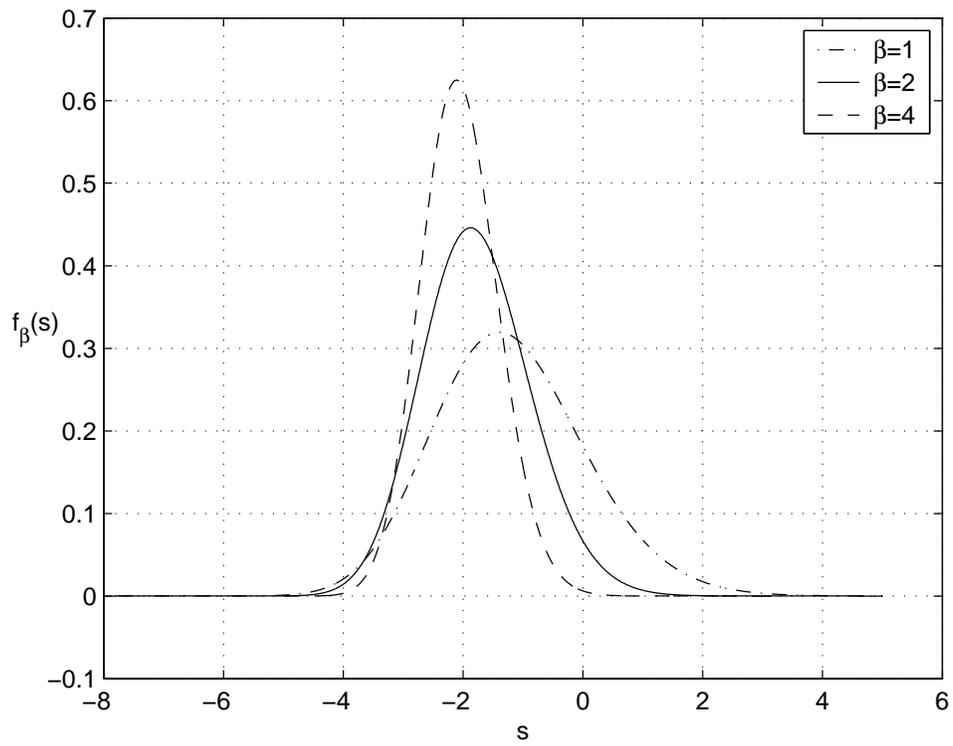}
  \end{center}
  \caption{The probability distributions $f_1(s)$, $f_2(s)$, and $f_4(s)$,
computed using the Painlev\'e II solution.}
  \label{fig:plt2}
\end{figure}

Finally, plot the curves:
\begin{small} \begin{verbatim}
  plot(s,f1,s,f2,s4,f4)
  legend('\beta=1','\beta=2','\beta=4')
  xlabel('s')
  ylabel('f_\beta(s)','rotation',0)
  grid on
\end{verbatim} \end{small}
The result can be seen in Figure~\ref{fig:plt2}.

\clearpage
\section{Eigenvalue Spacings Distributions}

Another quantity with an interesting probability distribution is the
spacings of the eigenvalues of random matrices. It turns out that the
eigenvalues are almost uniformly distributed, which means that every
random matrix gives a large number of spacings. The distributions can
then be efficiently computed by simulation.

Two other methods are used to compute the spacings distribution --
the solution of the Painlev\'e V nonlinear differential equation
and the eigenvalues of the Prolate matrix. Finally, the results are
compared with the spacings of the zeros along the critical line of
the Riemann zeta function.

\subsection{Simulation}

As before, the simulations are made with matrices from the Gaussian
Unitary Ensemble. The normalized spacings of the eigenvalues
$\lambda_1\le\lambda_2\le\ldots\le\lambda_N$ are computed according to
\begin{align}
\delta'_k=\frac{\lambda_{k+1}-\lambda_k}{\pi\beta}\sqrt{2\beta n-\lambda_k^2},
 \qquad k\approx n/2,
\end{align}
where $\beta=2$ for the GUE. The distribution of the eigenvalues is
almost uniform, with a slight deviation at the two ends of the
spectrum. Therefore, only half of the eigenvalues are used, and one
quarter of the eigenvalues at each end is discarded.

Again, to allow for a more efficient simulation, the tridiagonal matrix
(\ref{eq:Hbeta}) is used instead of the full Hermitian matrix. In this case,
all the eigenvalues are computed, which can be done in a time proportional
to $n^2$. While this could in principle be done using the MATLAB sparse
matrix structure and the \texttt{eigs} function, the more efficient
\texttt{trideig} function is used below to compute all the eigenvalues of
a symmetric tridiagonal matrix. It can be downloaded from
\textsf{http://www-math.mit.edu/$\sim$persson/mltrid}.

\begin{figure}
  \begin{center}
    \includegraphics[width=0.8\textwidth]{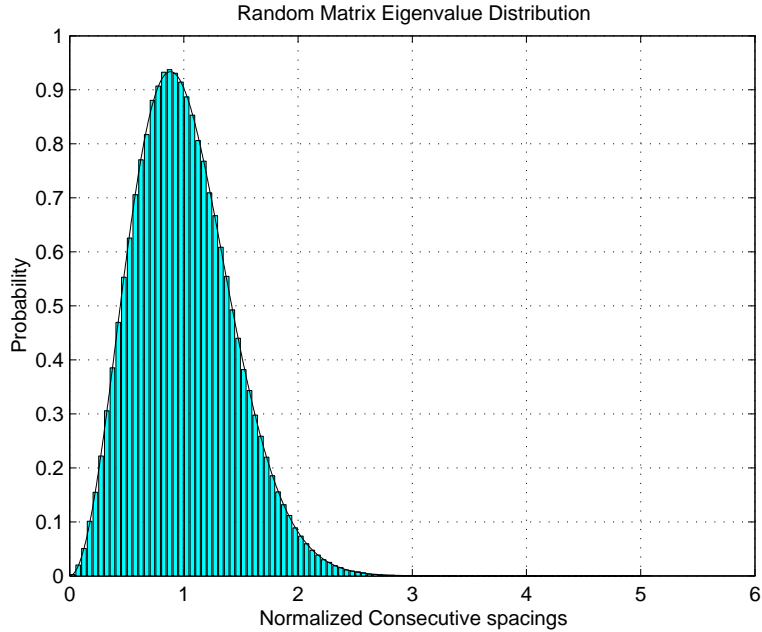}
  \end{center}
  \caption{Probability distribution of consecutive spacings of
random matrix eigenvalues ($1000$ repetitions, $n=1000$)}
  \label{fig:plt3}
\end{figure}

The histogram can now be computed by simulation with the following lines
of code. Note that the function \texttt{chi2rnd} from the Statistics
Toolbox is required.
\begin{small} \begin{verbatim}
  n=1000;
  nrep=1000;
  beta=2;
  
  ds=zeros(1,nrep*n/2);
  for ii=1:nrep
    l=trideig(randn(n,1),sqrt(chi2rnd((n-1:-1:1)'*beta)/2));
    d=diff(l(n/4:3*n/4))/beta/pi.*sqrt(2*beta*n-l(n/4:3*n/4-1).^2);
    ds((ii-1)*n/2+1:ii*n/2)=d;
  end
  
  histdistr(ds,0:0.05:5);
\end{verbatim} \end{small}
The resulting histogram can be found in Figure~\ref{fig:plt3}. The figure
also shows the expected curve as a solid line.

\subsection{Painlev\'e V}

\begin{figure}
  \begin{center}
    \includegraphics[width=1\textwidth]{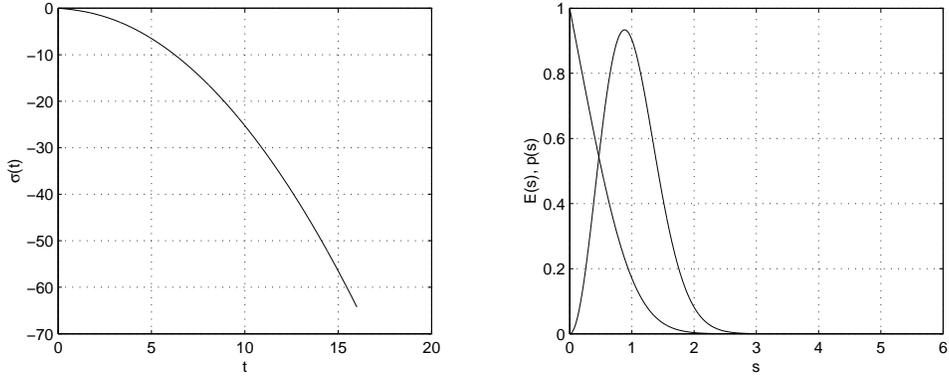}
  \end{center}
  \caption{Painlev\'e V (left), $E(s)$ and $p(s)$ (right).}
  \label{fig:plt4}
\end{figure}

The probability distribution $p(s)$ for the eigenvalue spacings when
$\beta=2$ can be computed with the solution to the Painlev\'e V
nonlinear differential equation (see \cite{TracyWidomIntroduction} for details):
\begin{align} \label{eq:PainleveV}
(t\sigma'')^2+4(t\sigma'-\sigma)\left(t\sigma'-\sigma+(\sigma')^2\right)=0
\end{align}
with the boundary condition
\begin{align} \label{eq:PainleveVbnd}
\sigma(t)\approx -\frac{t}{\pi}-\left(\frac{t}{\pi}\right)^2,
   \qquad\textrm{as }t\rightarrow 0^+.
\end{align}
Then $p(s)$ is given by
\begin{align}
p(s)=\frac{d^2}{ds^2}E(s)
\end{align}
where
\begin{align} \label{eq:Es}
E(s)=\exp \left( \int_0^{\pi s} \frac{\sigma(t)}{t} dt \right).
\end{align}
Explicit differentiation gives
\begin{align}
p(s)=\frac{1}{s^2}
\left(
\pi s \sigma'(\pi s)-\sigma(\pi s)+\sigma(\pi s)^2
\right)
E(s).
\end{align}
The second-order differential equation (\ref{eq:PainleveV}) can be
written as a first-order system of differential equations:
\begin{align} \label{eq:PainleveVsystem}
\frac{d}{dt}
\begin{pmatrix}
\sigma \\ \sigma'
\end{pmatrix}
=
\begin{pmatrix}
\sigma' \\ -\frac{2}{t}\sqrt{(\sigma-t\sigma')\left(t\sigma'-\sigma+(\sigma')^2\right)}
\end{pmatrix}.
\end{align}
This is solved as an initial-value problem starting at $t=t_0=\,$very
small positive number. The value $t=0$ has to be avoided because of
the division by $t$ in the system of equations. This is not a
problem, since the boundary condition (\ref{eq:PainleveVbnd}) provides
an accurate value for $\sigma(t_0)$ (as well as $E(t_0/\pi)$). The
boundary conditions for the system (\ref{eq:PainleveVsystem}) then
become
\begin{align}
\begin{cases}
\sigma(t_0) &= -\frac{t_0}{\pi}-\left(\frac{t_0}{\pi}\right)^2 \\
\sigma'(t_0) &= -\frac{1}{\pi}-\frac{2t_0}{\pi}.
\end{cases}
\end{align}
To be able to compute $E(s)$ using (\ref{eq:Es}), the variable
\begin{align}
I(t)=\int_0^t \frac{\sigma(t')}{t'}\,dt'
\end{align}
is added to the system, as well as the equation
$\frac{d}{dt}I=\frac{\sigma}{t}$. The corresponding initial value is
\begin{align}
I(t_0)\approx \int_0^{t_0} \left(-\frac{1}{\pi}-\frac{t}{\pi^2}\right)\,dt
=-\frac{t_0}{\pi}-\frac{t_0^2}{2\pi^2}.
\end{align}
Putting it all together, the final system is
\begin{align}
\frac{d}{dt}
\begin{pmatrix}
\sigma \\ \sigma' \\ I
\end{pmatrix}
=
\begin{pmatrix}
\sigma' \\ -\frac{2}{t}\sqrt{(\sigma-t\sigma')\left(t\sigma'-\sigma+(\sigma')^2\right)} \\
\frac{\sigma}{t}
\end{pmatrix}
\end{align}
with boundary condition
\begin{align}
\begin{pmatrix}
\sigma(t_0) \\ \sigma'(t_0) \\ I(t_0)
\end{pmatrix}
=
\begin{pmatrix}
-\frac{t_0}{\pi}-\left(\frac{t_0}{\pi}\right)^2 \\
-\frac{1}{\pi}-\frac{2t_0}{\pi} \\
-\frac{t_0}{\pi}-\frac{t_0^2}{2\pi^2}
\end{pmatrix}.
\end{align}
This system is defined as an inline function in MATLAB:
\begin{small} \begin{verbatim}
  desig=inline(['[y(2); -2/t*sqrt((y(1)-t*y(2))*' ...
                '(t*y(2)-y(1)+y(2)^2)); y(1)/t]'],'t','y');
\end{verbatim} \end{small}
Specify the integration interval and the desired output times:
\begin{small} \begin{verbatim}
  t0=1e-12;
  tn=16;
  tspan=linspace(t0,tn,1000);
\end{verbatim} \end{small}
Set the initial condition:
\begin{small} \begin{verbatim}
  y0=[-t0/pi-(t0/pi)^2; -1/pi-2*t0/pi; -t0/pi-t0^2/2/pi^2];
\end{verbatim} \end{small}
Finally, set the integration tolerances and call the solver:
\begin{small} \begin{verbatim}
  opts=odeset('reltol',1e-13,'abstol',1e-14);
  [t,y]=ode45(desig,tspan,y0,opts);
\end{verbatim} \end{small}
The solution components are now in the columns of \texttt{y}. Use
these to evaluate $E(s)$ and $p(s)$:
\begin{small} \begin{verbatim}
  s=t/pi;
  E=exp(y(:,3));
  p=1./s.^2.*E.*(t.*y(:,2)-y(:,1)+y(:,1).^2);
  p(1)=2*s(1); % Fix due to cancellation
\end{verbatim} \end{small}
A plot of $p(s)$ can be made with the command
\begin{small} \begin{verbatim}
  plot(s,p)
  grid on
\end{verbatim} \end{small}
and it can be seen in Figure~\ref{fig:plt4}. Plots are also shown
of $E(s)$ and $\sigma(t)$.

\subsection{The Prolate Matrix}

Another method to calculate the distribution of the eigenvalue spacings
is to compute the eigenvalues $\lambda_i$ of the operator
\begin{align}
f(y) \rightarrow \int_{-1}^1 Q(x,y)f(y)\,dy, \qquad
Q(x,y)=\frac{\sin\left((x-y)\pi t\right)}{(x-y)\pi}.
\end{align}
Then $E(2t)=\prod_i (1-\lambda_i)$, and $p(s)$ can be computed as before.
To do this, first define the infinite symmetric Prolate matrix:
\begin{align}
A_\infty=
\begin{pmatrix}
a_0 & a_1 & \ldots \\
a_1 & a_0 & \ldots \\
\vdots & \vdots & \ddots
\end{pmatrix}
\end{align}
with $a_0=2w$, $a_k=(\sin 2\pi wk)/\pi k$ for $k=1, 2, \ldots$, and
$0<w<\frac{1}{2}$. A discretization of $Q(x,y)$ is achieved by setting
$w=t/n$ and extracting the upper-left $n\times n$ submatrix $A_n$ of $A_\infty$.

Below, the full matrix $A_n$ is used, and all the eigenvalues are
computed in $n^3$ time using the \texttt{eig} function. However, $A_n$
commutes with the following symmetric tridiagonal matrix \cite{Slepian},
and therefore has the same eigenvectors:
\begin{align}
T_n=
\begin{pmatrix}
\alpha_1 & \beta_1 & & & \\
\beta_1 & \alpha_2 & \beta_2 & & \\
 & \ddots & \ddots & \ddots & \\
 & & \beta_{n-2} & \alpha_{n-1} & \beta_{n-1} \\
 & & & \beta_{n-1} & \alpha_n
\end{pmatrix}
\end{align}
where
\begin{align}
\left\{
\begin{array}{rcl}
\alpha_k &=& \left(\frac{n+1}{2}-k\right)^2\cos 2\pi w \\
\beta_k &=& \frac{1}{2}k(n-k).
\end{array}
\right.
\end{align}
It is then in principle possible to use the new techniques described
in \cite{Dhillon} to compute all the eigenvalues and eigenvectors of
$T_n$ in $n^2$ time, and then get the eigenvalues of $A_n$ by dot
products. This is not done in this example.

The code for computing $E(s)$ is shown below. This time, $p(s)$ is
evaluated by numerical differentiation since no information about the
derivative of $E(s)$ is available.
\begin{small} \begin{verbatim}
  s=0:0.01:5;
  n=100;
  E0=zeros(size(s));
  for ii=1:length(s)
    Q=gallery('prolate',n,s(ii)/2/n);
    E0(ii)=prod(1-eig(Q));
  end
  p0=gradient(gradient(E0,s),s);
\end{verbatim} \end{small}
To improve the accuracy in $E(s)$, Richardson extrapolation can be
used. This is done as follows, where the values are assumed to
converge as $1/n^2$:

\begin{small} \begin{verbatim}
  % ... Compute s and E using Painleve V in previous section
  
  Es=zeros(length(t),0);
  E1=zeros(size(s));
  for n=20*2.^(0:3)
    for ii=1:length(s)
      Q=gallery('prolate',n,s(ii)/2/n);
      E1(ii)=prod(1-eig(Q));
    end
    Es=[Es,E1];
  end
  
  for ii=1:3
    max(abs(Es-E(:,ones(1,size(Es,2)))))
    Es=Es(:,2:end)+diff(Es,1,2)/(2^(ii+1)-1);
  end
  max(abs(Es-E))
\end{verbatim} \end{small}
The errors $\max_{0\le s\le 5} \left| E_1(s)-E(s) \right|$ are shown
in Table 1, for $n=20,40,80$, and $160$. The error after all
extrapolations is of the same order as the ``exact solution'' using
Painlev\'e V.

\begin{table} \label{tab1}
\begin{center}
\begin{tabular}{|r|r|r|r|r|}
\hline N & Error 0 & Error 1 & Error 2 & Error 3 \\ \hline 20 & 0.2244
& & & \\ 40 & 0.0561 & 0.7701 & & \\ 80 & 0.0140 & 0.0483 & 0.5486 &
\\ 160 & 0.0035 & 0.0032 & 0.0323 & 2.2673 \\ \hline & $\cdot 10^{-3}$
& $\cdot 10^{-7}$ & $\cdot 10^{-8}$ & $\cdot 10^{-11}$ \\ \hline
\end{tabular}
\end{center}
\caption{Difference between Prolate solution $E_1(s)$ and Painlev\'e V solution $E(s)$
after 0, 1, 2, and 3 Richardson extrapolations.}
\end{table}

\subsection{Riemann Zeta Zeros}

It has been observed that the zeros of the Riemann zeta function along
the critical line $\mathrm{Re}(z)=\frac{1}{2}$ behave similar to the
eigenvalues of random matrices in the GUE. Here, the distribution of
the scaled spacings of the zeros is compared to the corresponding
distribution for eigenvalues computed using the Painlev\'e V equation
from the previous chapters.

\begin{figure}
  \begin{center}
    \includegraphics[width=0.8\textwidth]{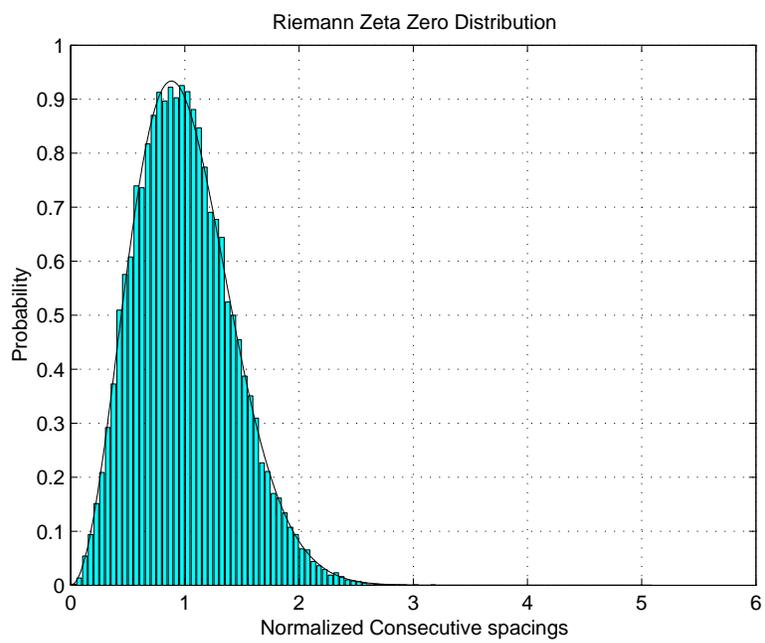}
  \end{center}
  \caption{Probability distribution of consecutive spacings of
Riemann zeta zeros ($30,000$ zeros, $n\approx 10^{12}, 10^{21}, 10^{22}$)}
  \label{fig:plt5}
\end{figure}

Define the $n$th zero $\gamma_n=n^\mathrm{th}$ as
\begin{align}
\zeta\left(\frac{1}{2}+i\gamma_n\right) = 0,\qquad 0<\gamma_1<\gamma_2<\ldots
\end{align}
Compute a normalized spacing:
\begin{align}
\tilde{\gamma}_n=\frac{\gamma_n}{\textrm{av spacing near $\gamma_n$}}=
\gamma_n\cdot \left[\frac{\log \gamma_n/2\pi}{2\pi}\right].
\end{align}
Zeros of the Riemann zeta function can be downloaded from \cite{Odlyzko}.
Assuming that the MATLAB variable \texttt{gamma} contains the zeros,
and the variable \texttt{offset} the offset, these two lines compute
the consecutive spacings $\tilde{\gamma}_{n+1}-\tilde{\gamma}_n$ and
plot the histogram:
\begin{small} \begin{verbatim}
  delta=diff(gamma)/2/pi.*log((gamma(1:end-1)+offset)/2/pi);
  histdistr(delta,0:0.05:5.0);
\end{verbatim} \end{small}
The result can be found in Figure~\ref{fig:plt5}, along with the
Painlev\'e V distribution. The curves are indeed in good agreement,
although the number of samples here is a little to low to get a
perfect match.

\clearpage
\bibliographystyle{plain}
\bibliography{numrand}

\end{document}